\documentclass[twocolumn,aps,prd,preprintnumbers,showpacs,superscriptaddress,nofootinbib,amsmath,amssymb,floats,floatfix,showkeys,notitlepage,longbibliography]{revtex4-1}

\usepackage{orcidlink}
\usepackage{comment}
\usepackage{lipsum}
\usepackage{graphicx}
\usepackage{subfigure}
\usepackage{palatino}
\usepackage{sans}
\usepackage{hyperref}
\hypersetup{colorlinks=true,linkcolor=blue,urlcolor=blue,citecolor=blue}
\usepackage[toc,page]{appendix}
\usepackage[normalem]{ulem}
\usepackage{adjustbox}
\usepackage{latexsym}
\usepackage{amsmath}
\usepackage{amssymb}
\usepackage{amsfonts}
\usepackage{dcolumn}
\usepackage{bm}
\usepackage{tikz}
\usepackage{bigints}
\usepackage{array,tabularx,multirow,booktabs}
\usepackage[tracking=true]{microtype}
\usepackage{soul} 
\usepackage{booktabs}
\usepackage{multirow}
\SetTracking{}{500}
\SetTracking{encoding={*}, shape=sc}{40}
\UseRawInputEncoding 
\allowdisplaybreaks

\begin{document} \sloppy
\title{Phenomenology of Schwarzschild-like Black Holes with a Generalized Compton Wavelength}


\author{Reggie C. Pantig \orcidlink{0000-0002-3101-8591}} 
\email{rcpantig@mapua.edu.ph}
\affiliation{Physics Department, Map\'ua University, 658 Muralla St., Intramuros, Manila 1002, Philippines.}

\author{Ali \"Ovg\"un \orcidlink{0000-0002-9889-342X}}
\email{ali.ovgun@emu.edu.tr}
\affiliation{Physics Department, Eastern Mediterranean University, Famagusta, 99628 North
Cyprus via Mersin 10, Turkiye.}

\author{Gaetano Lambiase \orcidlink{0000-0001-7574-2330}}
\email{lambiase@sa.infn.it}
\affiliation{Dipartimento di Fisica ``E.R Caianiello'', Università degli Studi di Salerno, Via Giovanni Paolo II, 132 - 84084 Fisciano (SA), Italy.}
\affiliation{Istituto Nazionale di Fisica Nucleare - Gruppo Collegato di Salerno - Sezione di Napoli, Via Giovanni Paolo II, 132 - 84084 Fisciano (SA), Italy.}

\begin{abstract}
We investigate the influence of the generalized Compton wavelength (GCW), emerging from a three-dimensional dynamical quantum vacuum (3D DQV) on Schwarzschild-like black hole spacetimes, motivated by the work of Fiscaletti [10.1134/S0040577925020096] \cite{Fiscaletti:2025iuh}. The GCW modifies the classical geometry through a deformation parameter $ \varepsilon $, encoding quantum gravitational backreaction. We derive exact analytical expressions for the black hole shadow radius, photon sphere, and weak deflection angle, incorporating higher-order corrections and finite-distance effects of a black hole with generalized Compton effect (BHGCE). Using Event Horizon Telescope (EHT) data, constraints on $ \varepsilon $ are obtained: $ \varepsilon \in [-2.572, 0.336] $ for Sgr. A* and $ \varepsilon \in [-2.070, 0.620] $ for M87*, both consistent with general relativity yet allowing moderate deviations. Weak lensing analyses via the Keeton-Petters and Gauss-Bonnet formalisms further constrain $ \varepsilon \approx 0.061 $, aligning with solar system bounds. We compute the modified Hawking temperature, showing that positive $ \varepsilon $ suppresses black hole evaporation. Quasinormal mode frequencies in the eikonal limit are also derived, demonstrating that both the oscillation frequency and damping rate shift under GCW-induced corrections. Additionally, the gravitational redshift and scalar perturbation waveform exhibit deformations sensitive to $ \varepsilon $. Our results highlight the GCW framework as a phenomenologically viable semiclassical model, offering testable predictions for upcoming gravitational wave and VLBI observations.
\end{abstract}


\date{\today}
\maketitle

\section{Introduction} \label{intro}
The union between quantum mechanics and gravity remains one of the deepest frontiers in theoretical physics. Central to this effort is the quest for a unified framework that reconciles quantum principles with spacetime curvature. One promising path emerges through the Generalized Uncertainty Principle (GUP)--a quantum gravity-inspired extension of Heisenberg's principle--which posits a minimal length scale and predicts modifications to both quantum and gravitational phenomena at the Planck scale. Foundational work in this area, particularly by Scardigli \cite{Scardigli_1999}, has shown that GUP arises naturally in thought experiments involving micro black holes, establishing the theoretical groundwork for phenomenological extensions of general relativity.

Building on these insights, researchers such as Carr and Lake have proposed the Compton-Schwarzschild correspondence, a duality between the Compton wavelength and Schwarzschild radius that becomes evident when quantum gravitational effects are considered. This correspondence implies the existence of sub-Planckian black holes--entities with mass below the Planck scale but radii comparable to their Compton wavelength \cite{Carr:2011pr,Carr_2015,Carr_2015(2)}. These results have been reinforced by modified wave-packet descriptions and extended de Broglie relations that interpolate between quantum and relativistic limits \cite{Lake_2015}.

The broader implications of these frameworks are manifold. Ref. \cite{Carr_2022} explores how extra dimensions modify the duality between quantum and gravitational length scales, impacting both particle physics and black hole models. Complementary studies investigate how GUP modifies black hole thermodynamics, entropy bounds, and even the Wheeler-DeWitt equation, suggesting novel behavior near singularities \cite{Bina_2010,Vayenas_2015}. Recent work has even drawn connections between GUP and quantum information theory, where black hole entropy may reflect underlying entanglement structures \cite{daSilva:2022xgx}. The phenomenological properties of hypothetical “GUP stars,” compact objects governed by the generalized uncertainty principle, were analyzed \cite{Buoninfante:2020cqz}. Connections between a nonzero GUP parameter and Lorentz‑violating terms in effective field theories were investigated \cite{Scardigli:2018lbm}. A value for the GUP deformation parameter was derived by computing quantum corrections to the Newtonian gravitational potential \cite{Scardigli:2016pjs}. Coherent states satisfying generalized uncertainty relations were constructed and shown to act as Tsallis‑type probability amplitudes in nonextensive thermostatistics \cite{Jizba:2023ygi}. The relationship between the GUP framework and asymptotically safe gravity scenarios was examined \cite{Lambiase:2022xde}. A mechanism by which the generalized uncertainty principle could generate the observed baryon asymmetry in the early Universe was proposed \cite{Das:2021nbq}. Implications of the generalized uncertainty principle for three‑dimensional gravity and the BTZ black hole were analyzed \cite{Iorio:2019wtn}. Interplay between noncommutative Schwarzschild geometry and GUP‑induced modifications was studied \cite{Kanazawa:2019llj}. Emergence of Lorentz‑violating operators alongside a generalized uncertainty principle in effective quantum gravity models was explored \cite{Lambiase:2017adh}. The quantum‑corrected scattering cross‑section of a Schwarzschild black hole incorporating GUP effects was computed \cite{Heidari:2023ssx}. EUP‑corrected thermodynamic properties of the BTZ black hole were derived \cite{Hamil:2022bpd}. Thermodynamics of a Schwarzschild black hole surrounded by quintessence under GUP was analyzed \cite{Lutfuoglu:2021ofc}. Thermodynamic behavior of Schwarzschild and Reissner-Nordstr\"om black holes under a higher‑order GUP was investigated \cite{Hassanabadi:2021kuc}. Generalized Klein–Gordon oscillator dynamics in a cosmic spacetime with a space‑like dislocation and associated Aharonov–Bohm effect were studied \cite{Lutfuoglu:2020wjy}. Applications of a new higher‑order generalized uncertainty principle to quantum and gravitational systems were explored \cite{Hamil:2020ldl}.

In parallel, observationally relevant signatures of these quantum-corrected black holes--such as black hole shadows and weak deflection angles--have become focal points in recent research. As observed by the Event Horizon Telescope, the shape and size of black hole shadows are sensitive to spacetime modifications \cite{Zakharov:2023yjl,Zakharov:2021gbg,Zakharov:2014lqa,Khodadi:2020jij,Vagnozzi:2022moj,Vagnozzi:2019apd,Allahyari:2019jqz}. Neves and collaborators established an upper bound on the generalized uncertainty principle parameter by analyzing its effects on the shadow of a Schwarzschild black hole \cite{Neves:2019lio}.  Tamburini, Feleppa, and Thide derived constraints on the GUP by studying the twisting of light by rotating black holes and comparing with the Event Horizon Telescope image of M87* \cite{Tamburini:2021inp}. Anacleto et al. investigated how quasinormal mode frequencies and the apparent shadow of a Schwarzschild black hole are modified under a GUP framework \cite{Anacleto:2021qoe}. Karmakar et al. examined the thermodynamics and shadow structure of GUP-corrected black holes with topological defects within the context of Bumblebee gravity \cite{Karmakar:2023mhs}. Ong critiqued aspects of the effective metric derived from the generalized uncertainty principle, highlighting conceptual and technical issues in previous formulations \cite{Ong:2023jkp}. Lambiase et al. explored connections between the GUP and asymptotically safe gravity by analyzing black hole shadows and quasinormal modes \cite{Lambiase:2023hng}. Hoshimov and colleagues studied weak gravitational lensing and the shadow of a GUP-modified Schwarzschild black hole in the presence of a plasma environment \cite{Hoshimov:2023tlz}. Chen et al. assessed thermodynamic properties, evaporation times, and shadow constraints of GUP-corrected black holes using EHT observations of M87* and Sgr A* \cite{Chen:2024mlr}. One introduced a generalized extended uncertainty principle to analyze black hole shadows and lensing effects across macro- and microscopic scales \cite{Lobos:2022jsz}. Moreover, it has been applied Event Horizon Telescope data to constrain black hole solutions influenced by dark matter under a GUP minimal length-scale effect \cite{Ovgun:2023wmc}. Chaudhary et al. investigated the imaging signatures and stability of black holes surrounded by a cloud of strings and quintessence within an extended GUP framework \cite{Chaudhary:2024yod}. Similarly, AGEUP-inspired metrics incorporating spacetime curvature effects yield predictions for deflection angles and shadows that could be tested with future VLBI observations \cite{Pantig_2025}. Further developments have emerged from nonlocal gravity frameworks \cite{Fu_2022}, loop quantum gravity corrections \cite{Alloqulov_2025}, and even quantum fuzziness models based on multifractional geometries \cite{Pantig:2024fbh}.

In this evolving context, the work in Ref. \cite{Fiscaletti:2025iuh} introduces a compelling and underexplored approach. His formulation--based on a three-dimensional dynamical quantum vacuum (3D DQV)--proposes a generalized Compton wavelength derived from vacuum energy density fluctuations, leading to modified expressions for black hole metrics and event horizons. While rich in conceptual innovation, this model has not yet been applied to observational features such as shadows and deflection angles, nor has it been framed within the broader body of GUP-based and phenomenological gravity research. The present study aims to bridge this gap: by extending Fiscaletti's results to compute the black hole shadow radius and weak deflection angle, we explore whether his quantum vacuum framework yields distinctive, testable predictions that align with--or deviate from--other GUP-based theories.

The paper is organized as follows: In Section \ref{sec2}, we provide a brief overview of quantum-geometric corrections to general relativity via the generalized Compton wavelength (GCW) formalism, which yields a quantum-modified Schwarzschild-like metric. Section \ref{sec3} explores the shadow cast by this modified black hole, leading to analytical expressions for the photon sphere and shadow radius, with constraints on the quantum deformation parameter $ \varepsilon $ derived from Event Horizon Telescope (EHT) observations. In Section \ref{sec:kp-deflection}, we apply the Keeton-Petters formalism to compute the weak deflection angle and its impact on lensing observables, including image magnification and time delays. Section \ref{sec5} complements this with an independent calculation via the Gauss-Bonnet theorem, extending the deflection angle analysis to finite distances and including massive particles. Section \ref{sec:hawking-temperature} addresses the modified Hawking temperature arising from the deformed lapse function and shows how $ \varepsilon $ modifies black hole thermodynamics. In Section \ref{sec7}, we study the quasinormal mode (QNM) spectrum in the eikonal limit, relating the damping and oscillation frequencies to the photon sphere’s location and stability. Section \ref{sec8} derives the gravitational redshift for static observers near the black hole, while Section \ref{sec9} numerically analyzes scalar perturbations and ringdown waveforms, revealing the influence of quantum corrections on echo structures and late-time tails. Finally, unless otherwise specified, we used $G=c=1$, metric signature of $(-,+,+,+)$.

\section{Brief Review of Quantum-Geometric Corrections to General Relativistic Effects via Generalized Compton Wavelength Formalism} \label{sec2}
In recent explorations aimed at reconciling quantum mechanics with general relativity, particular attention has been devoted to the formulation of generalized uncertainty principles (GUP) and their ramifications for both microphysical and macrophysical regimes. A compelling development in this direction is offered by Fiscaletti \cite{Fiscaletti:2025iuh}, who proposes a unifying framework predicated upon a generalized Compton wavelength derived from energy-density fluctuations in a three-dimensional dynamical quantum vacuum (3D DQV). This approach yields a quantum-modified extension of the Schwarzschild geometry and engenders corrections to canonical predictions of general relativity, most notably light deflection, perihelion precession, gravitational redshift, and time dilation.

The theoretical starting point is a modified uncertainty relation, incorporating a deformation parameter $ \beta $ and the vacuum's variable quantum energy density $ \Delta\rho_{qvE} $:
\begin{equation}
\Delta x \Delta p \gtrsim \frac{\hbar}{2} \left( 1 + \beta \frac{l_P^2 \Delta\rho_{qvE}^2 V^2}{\hbar^2 c^2} \right).
\end{equation}
This relation implies the existence of a minimal measurable length, especially pertinent at Planckian scales. The associated fluctuation in vacuum energy density for a particle of mass $ m $ within volume $ V $ is defined as:
\begin{equation}
\Delta\rho_{qvE} = \rho_{PE} - \rho_{qvE} = \frac{mc^2}{V}, \quad \text{where} \quad \rho_{PE} = \frac{M_P c^2}{l_P^3}.
\end{equation}

From these premises, a generalized Compton wavelength (GCW) is derived that smoothly interpolates between the Compton scale of particles and the Schwarzschild radius of black holes:
\begin{equation} \label{e3}
R'_C = R'_S = \sqrt{ \left( \frac{\beta \hbar c}{\Delta\rho_{qvE} V} \right)^2 + \left( \frac{\beta l_P^2 \Delta\rho_{qvE} V}{ \hbar c} \right)^2 }.
\end{equation}
Eq. \eqref{e3} embodies the central unifying principle of the model: that micro- and macro-scale entities are manifestations of the same quantum-geometric substrate, governed by vacuum fluctuations and encoded in the GCW.

The GCW is then employed to deform the Schwarzschild metric $ds^2 = -F(r) dt^2 + F(r)^{-1} dr^2 + C(r) d\phi^2$, where $C(r) = r^2$ and $\theta = \pi/2$, yielding a quantum-modified lapse function \cite{Fiscaletti:2025iuh}:
\begin{align} \label{e4}
F(r) &=\; 1 - \frac{1}{r} \sqrt{ 
\left( \frac{\beta \hbar c}{\Delta\rho_{qvE} V} \right)^2 
+ \left( \frac{\beta l_P^2 \Delta\rho_{qvE} V}{ \hbar c} \right)^2 } \nonumber \\
& + \frac{\varepsilon}{4r^2} \left[ 
\left( \frac{\beta \hbar c}{\Delta\rho_{qvE} V} \right)^2 
+ \left( \frac{\beta l_P^2 \Delta\rho_{qvE} V}{ \hbar c} \right)^2 
\right].
\end{align}

Here, $ \varepsilon $ is a dimensionless deformation parameter characterizing quantum backreaction. Within this framework, even the mass of a black hole becomes an emergent quantity, expressed in terms of the GCW:
\begin{equation} \label{e5}
M = \frac{c^2 (1 + \sqrt{1 - \varepsilon})^3}{8 G (1 - \varepsilon + \sqrt{1 - \varepsilon})} \sqrt{ \left( \frac{\beta \hbar c}{\Delta\rho_{qvE} V} \right)^2 + \left( \frac{\beta l_P^2 \Delta\rho_{qvE} V}{ \hbar c} \right)^2 }.
\end{equation}
This reformulation signifies a deep ontological shift: mass itself is no longer a primary attribute but arises from quantum fluctuations of vacuum energy within the 3D DQV framework.

Substituting Eq. \eqref{e5} to Eq. \eqref{e4}, we can effectively recast the lapse function as
\begin{align}
    F(r) &= 1 - \frac{2M}{r} \left[ \frac{4\left(1 - \varepsilon + \sqrt{1 - \varepsilon}\right)}{\left(1 + \sqrt{1 - \varepsilon}\right)^3} \right] \nonumber \\
    & + \frac{16\varepsilon M^2}{r^2} \left[\frac{\left(1 - \varepsilon + \sqrt{1 - \varepsilon}\right)}{\left(1 + \sqrt{1 - \varepsilon}\right)^3}\right]^2,
\end{align}
which shows dimensional consistency. It can be seen that the domain of $\varepsilon$ is $(-\infty,1]$. With a little deviation from the standard Schwarzschild geometry, where $\varepsilon \rightarrow 0$, we can write

We can further simplify it by introducing
\begin{equation}
    \Sigma = \frac{4\left(1- \varepsilon + \sqrt{1-\varepsilon}\right)}{\left(1+\sqrt{1-\varepsilon}\right)^{3}},
\end{equation}
yielding
\begin{equation} \label{e11}
    F(r) = 1 - \frac{2M \Sigma}{r} + \frac{\varepsilon M^2 \Sigma^2}{r^2}.
\end{equation}
For the succeeding sections, we assume a black hole metric that is static and spherically symmetric, and specialize only along the equatorial plane where $\theta = \pi/2$. Then we have a black hole metric with $1+2$ dimensionality:
\begin{equation} \label{e9}
    ds^{2} = -F(r) dt^{2} + F(r)^{-1} dr^{2} + r^2 d\phi^{2}.
\end{equation}

\section{Shadow of the BHGCE} \label{sec3}
The study of black hole shadows has emerged as a powerful probe of strong-field gravity, particularly in testing deviations from the classical Schwarzschild and Kerr metrics. With the advent of the Event Horizon Telescope (EHT) and its imaging of the supermassive black hole Sgr. A* and M87*, the size and shape of the shadow cast by a black hole can now be constrained with increasing precision. Within this observational context, the shadow radius encodes rich information about the underlying spacetime geometry. As such, black hole shadow observations offer a compelling means to place bounds on theoretical modifications to general relativity, including those induced by quantum gravitational corrections, such as those predicted by the generalized Compton wavelength (GCW) framework.

With Eq. \eqref{e11}, it is now possible to find the photon sphere radius analytically \cite{Claudel:2000yi}. The solution with physical significance is
\begin{equation} \label{eq_rps}
    r_{\rm ps} = \frac{M \Sigma}{2} \left( 3+\sqrt{-8 \varepsilon +9} \right)
\end{equation}
Using Eq. \eqref{eq_rps}, the critical impact parameter is found as
\begin{equation}
    b_{\rm crit}^2 = \frac{\left(3+\sqrt{-8 \varepsilon +9}\right)^{4} M^{2} \Sigma^{2}}{24+8 \sqrt{-8 \varepsilon +9}-16 \varepsilon}
\end{equation}
Finally, the exact expression for the shadow radius, with dependence on the observer's distance $r_{\rm obs}$, is found as
\begin{align} \label{e15}
    R_{\rm sh} = b_{\rm crit}\sqrt{A(r_{\rm obs})} &= \frac{\sqrt{2} M \Sigma \left( 3+\sqrt{-8 \varepsilon +9} \right)^2}{4 \left( 3+\sqrt{-8 \varepsilon +9}-2 \varepsilon \right)^{1/2}} \times \nonumber \\ 
    &\left( 1 - \frac{2M \Sigma}{r_{\rm obs}} + \frac{\varepsilon M^2 \Sigma^2}{r_{\rm obs}^2} \right)^{1/2}.
\end{align}

Let $\Delta_{EHT}$ represent the amount of uncertainty from the Schwarzschild radius found by the EHT. For Sgr. A* at $2\sigma$ level, $4.209M \leq R_{\rm Schw} \leq 5.560M$ which gives $\Delta_{EHT} \in [-0.364, 0.987]$ at $2\sigma$ level \cite{Vagnozzi:2022moj}, while for M87* at $1\sigma$ level, $4.313M \leq R_{\rm Schw} \leq 6.079M$ giving $\Delta_{EHT} \in [-0.883, 0.883]$ \cite{EventHorizonTelescope:2021dqv}. Then, equating Eq. \eqref{e15} to $R_{\rm Schw} \pm \Delta_{EHT}$, it is possible to find some numerical constraint for $\varepsilon$. For Sgr. A*, the constraint ranges to $\varepsilon \in [-2.572, 0.336]$, while for M87* $ \varepsilon \in [-2.070,0.620]$.

It can be seen that both bounds are now within a regime where the square roots appearing in the metric functions remain real, provided one adopts the admissible domain $ \varepsilon < 1 $. While negative values of $ \varepsilon $ remain allowed, the emergence of $ \varepsilon < -2.572 $ for Sgr. A* suggests that strong quantum corrections cannot yet be entirely excluded, though the plausibility of such large deformations should be weighed against additional constraints, e.g., from light deflection or redshift data. 

Crucially, the fact that $ \varepsilon = 0 $ (the Schwarzschild case) remains comfortably within both intervals implies that current EHT data are consistent with general relativity, but do not yet preclude moderate quantum gravitational corrections of the form introduced in this GCW-inspired framework. Future high-resolution shadow observations, especially those from next-generation VLBI arrays, may further narrow these bounds and provide sharper tests of the quantum nature of spacetime near black holes.

\section{Weak Deflection Angle of BHGCE via the Keeton-Petters Formalism}
\label{sec:kp-deflection}

A powerful observational tool to test gravitational theories beyond General Relativity (GR) is the measurement of gravitational lensing effects, especially the weak deflection angle of photons passing near massive astrophysical bodies. Modifications to GR can alter the spacetime geometry around massive compact objects, producing deviations in gravitational lensing signatures such as deflection angles, magnifications, image separations, and time delays. Precise measurements of these observables from astrophysical sources like galaxies, quasars, or supermassive black holes thus provide unique windows to constrain alternative gravitational theories.

Testing gravity theories beyond general relativity (GR) requires analyzing modifications to fundamental predictions, such as the gravitational deflection of light. The parametrized post-Newtonian (PPN) formalism provides a systematic approach to quantify deviations from GR through parameters characterizing modifications to the metric \cite{Epstein:1980dw,Will:2014kxa}.

In this section, we apply the Keeton-Petters formalism \cite{Keeton:2005jd,Ruggiero:2016iaq,Kumaran:2022soh}, a robust and systematic parametrized post-post-Newtonian (PPN) approach, to the static and spherically symmetric spacetime described by the line element presented in Eq. \eqref{e9}, with a metric function given by Eq. \eqref{e11}.

\subsection{Keeton-Petters formalism for weak deflection angle of BHGCE}

The Keeton-Petters formalism \cite{Keeton:2005jd} expresses the metric components in a series expansion around the Newtonian gravitational potential \(\phi=-\Sigma M/r\):
\begin{align}
A(r) &= 1 + 2 a_{1}\phi + 2 a_{2}\phi^{2} + \dots\,,\\[1ex]
B(r) &= 1 - 2 b_{1}\phi + 4 b_{2}\phi^{2} + \dots\,,
\end{align}
where \(A(r)\equiv f(r)\) and \(B(r)\equiv 1/f(r)\). Matching explicitly, we find the coefficients relevant to our spacetime are:
\begin{equation}
a_{1}=1,\quad b_{1}=1,\quad a_{2}=\frac{\varepsilon}{2},\quad b_{2}=\frac{(4-\varepsilon)}{4}.
\label{eq:coefficients}
\end{equation}

The weak gravitational deflection angle \(\hat{\alpha}\), up to second order in the gravitational radius \(M\) and impact parameter \(b\), is given by
\begin{equation}
\hat{\alpha} = A_{1}\frac{M}{b} + A_{2}\frac{M^2}{b^2} + \mathcal{O}\left(\frac{M^3}{b^3}\right),
\end{equation}
with coefficients
\begin{align}
A_{1}&=2(a_{1}+b_{1}),\\[1ex]
A_{2}&=\pi\left(2a_{1}^{2}-a_{2}+a_{1}b_{1}-\frac{b_{1}^{2}}{4}+b_{2}\right).
\end{align}

Substituting explicitly from Eq.~\eqref{eq:coefficients}, we find
\begin{align}
A_{1} &= 4\,,\\[1ex]
A_{2} &= \frac{\pi}{4}(11-3\varepsilon)\,,
\end{align}
thus yielding our explicit result for the deflection angle:
\begin{equation}
\hat{\alpha} = \frac{4M}{b} + \frac{\pi\,(11-3\varepsilon)}{4}\frac{M^2}{b^2}.
\label{eq:finaldeflection}
\end{equation}

It is particularly notable that for the Schwarzschild case (\(\varepsilon=0\)), Eq.~\eqref{eq:finaldeflection} reproduces exactly the well-known GR result \cite{Keeton:2005jd,Virbhadra:1999nm,Virbhadra:1998dy}.

\subsection{Implications for lensing observables of BHGCE}

In astrophysical gravitational lensing scenarios, the thin-lens approximation is commonly employed, where the deflecting mass is assumed to be compact and well localized. The lens equation connecting source and image positions (see e.g. Ref.~\cite{Ruggiero:2016iaq,Virbhadra:2024xpk,Kudo:2024aak,Virbhadra:2022iiy}) is written as:
\begin{equation}
D_{S} B = D_{S}\Theta - D_{LS}\hat{\alpha}\,,
\end{equation}
where \(D_L\), \(D_S\), and \(D_{LS}\) denote observer-lens, observer-source, and lens-source angular diameter distances, respectively, and \(\hat{\alpha}\) is the deflection angle given by Eq.~\eqref{eq:finaldeflection}. 

When the source aligns precisely behind the lens (\(B=0\)), one obtains the Einstein angular radius \(\theta_{E}\):
\begin{equation}
\theta_{E}=\sqrt{\frac{4MD_{LS}}{D_{L}D_{S}}}\,,
\end{equation}
a critical angular scale in gravitational lensing studies.

Higher-order expansions in the lens equation yield image positions \(\theta=\Theta/\theta_E\) as:
\begin{equation}
\theta=\theta_0+\frac{A_2}{A_1+4\theta_0^2}\epsilon,\quad\text{with}\quad\epsilon=\frac{\Theta_M}{\theta_E}=\frac{M}{D_{L}\theta_{E}}\ll1.
\end{equation}

For our spacetime, explicitly, this becomes:
\begin{equation}
\theta=\theta_0+\frac{\pi(11-3\varepsilon)}{16(1+\theta_0^2)}\,\epsilon.
\end{equation}

Magnifications \(\mu\) are particularly sensitive observables, given by
\begin{align}
\mu &= \mu_0 + \mu_1 \epsilon\,,\\[1ex]
\mu_0 &= \frac{\theta_0^4}{\theta_0^4 - 1}\,,\quad\quad
\mu_1 = -\frac{\pi\theta_0^3(11-3\varepsilon)}{(1+\theta_0^2)^3}\,,
\end{align}
which shows explicit dependency on the deviation parameters. Physically, deviations in magnifications from GR predictions can signal modified gravity effects or nonlinear electromagnetic backgrounds around lensing objects.

\subsection{Gravitational lensing time delay of BHGCE}

The gravitational lensing time delay \(\tau\), the difference in photon travel time due to spacetime curvature around the lensing mass, is one of the most precisely measurable lensing observables, particularly in strong-lensing systems of quasars and galaxies. Within the Keeton-Petters formalism, the time delay is expanded as:
\begin{equation}
\begin{split}
\frac{\tau}{\tau_E} = \frac{1}{2}\Biggl[\, & a_1 + \beta^2 - \theta_0^2 - \frac{a_1+b_1}{2}\ln\left(\frac{D_L\theta_0^2\theta_E^2}{4D_{LS}}\right) \,\Biggr] \\
&+ \frac{\pi}{16\theta_0}\left(8a_1^2 - 4a_2 + 4a_1b_1 - b_1^2 + 4b_2\right)\epsilon \\
&+ \mathcal{O}(\epsilon^2)\,.
\end{split}
\end{equation}

yielding explicitly for our metric:
\begin{equation}
\frac{\tau}{\tau_E}=\frac{1}{2}\left[1+\beta^2-\theta_0^2-\ln\left(\frac{D_L\theta_0^2\theta_E^2}{4D_{LS}}\right)\right]+\frac{3\pi(5-\varepsilon)}{16\theta_0}\epsilon\,.
\end{equation}

Therefore, the differential time delay between two observed images to first order becomes approximately:
\begin{equation}
\Delta\tau\,\epsilon\simeq\tau_E\frac{3\pi}{16}(5-\varepsilon)\epsilon\,,
\end{equation}
showing how gravitational lensing time delay measurements can explicitly constrain modifications to the spacetime geometry, particularly in future high-precision timing observations such as those achievable with next-generation space-based interferometry and pulsar timing experiments. These results emphasize the potential of gravitational lensing as a crucial probe into fundamental gravitational physics and modified gravity theories.

\section{Weak deflection angle of BHGCE via Gauss-Bonnet theorem method} \label{sec5}
In this section, we derive a more general expression for the weak deflection angle that includes the finite distance of the source S and the receiver R. Furthermore, we also include the deflection of massive particles, thereby implying the case of $v \neq 1$. For this purpose, we utilize the methods in Ref. \cite{Li:2020wvn}. We only show here the important steps.

We begin by considering the orbit equation of a test particle, derived from the geodesic equations in the deformed Schwarzschild geometry with metric coefficients modified by the GCW framework. The radial potential function $ F(u) $ governs the trajectory through
\begin{equation}
    F(u) = \frac{E^{2}}{J^{2}}-\left(1-2 M\Sigma u +\varepsilon M^2 \Sigma^2 u^2\right) \left(\frac{1}{J^{2}}+u^{2}\right),
\end{equation}
where $ u = 1/r $ is the reciprocal radial coordinate, $ E $ is the conserved energy, and $ J $ the angular momentum per unit mass of the particle. By employing a perturbative iterative method consistent with weak field conditions, an approximate solution to the orbit equation can be obtained:
\begin{equation}
    u(\phi) = \frac{\sin(\phi)}{b}+\frac{1+v^2\cos^2(\phi)}{b^2v^2}M\Sigma - \frac{\varepsilon M^2 \Sigma^2}{2v^2 b^3},
\end{equation}
where $ b = J/E $ is the impact parameter and $ v $ the velocity of the particle, which becomes unity in the massless (photon) case.

Next, the Gaussian curvature $ K $ of the Jacobi (optical) metric is integrated over a radial segment, with the upper limit determined by the perturbed geodesic path:
\begin{equation}
\begin{split}
\left[\int K\sqrt{g}\,dr\right]\bigg|_{r = r_\phi} =\; & -1 
+ \frac{\sin\left(\phi\right)\left(2E^2 - 1\right)M\Sigma}{\left(E^2 - 1\right)b} \\
& - \frac{3\varepsilon M^2 \Sigma^2 \sin^2\left(\phi\right)}{2b^2}.
\end{split}
\end{equation}
To obtain the total deflection, we now integrate the curvature over both the radial and angular extents of the geodesic region defined by the light path and the circular orbit at the photon sphere. This yields:
\begin{align}
&\int_{\phi_{\rm S}}^{\phi_{\rm R}} \int_{r_{\rm ph}}^{r(\phi)} K\sqrt{g} \, dr \, d\phi \;\sim\; 
 -\frac{(2E^2 - 1)M\Sigma}{(E^2 - 1)b} 
\cos\left(\frac{\phi}{\chi}\right)\bigg|_{\phi_{\rm S}}^{\phi_{\rm R}} \nonumber  \\
 &- \phi_{\rm RS} - \frac{3\varepsilon M^2 \Sigma^2}{4b^2}
\left(-\cos(\phi)\sin(\phi) + \phi\right)\bigg|_{\phi_{\rm S}}^{\phi_{\rm R}}.
\end{align}
To proceed, we must determine $ \phi $, the angular coordinate of the photon’s path, as a function of the impact parameter and the radial coordinate. By inverting the orbit solution and including quantum corrections, we obtain:
\begin{equation}
    \phi = \arcsin \! \left(b u \right) + \frac{\left[v^{2} \left(b^{2} u^{2}-1\right)-1\right] M\Sigma}{\sqrt{1-b^{2} u^{2}}\, b \,v^{2}} + \frac{\varepsilon M^2 \Sigma^2}{2 b^{2} \sqrt{1-b^{2} u^{2}}},
\end{equation}
which can be used to compute the angular separation between the source and receiver. Consequently, the cosine of the angle $ \phi $ can be expanded to include deformation corrections:
\begin{align}
    \cos(\phi) &= \sqrt{1-b^{2} u^{2}}-\frac{u \left(-1+v^{2} \left(b^{2} u^{2}-1\right)\right) M\Sigma}{\sqrt{1-b^{2} u^{2}}\, v^{2}} \nonumber \\
    &- \frac{u \,\varepsilon M^2 \Sigma^2}{2 b \sqrt{1-b^{2} u^{2}}},
\end{align}
providing the necessary ingredients to evaluate the curvature integrals explicitly. Finally, invoking the geometric symmetry of the setup and using the trigonometric identity
\begin{equation}
    \cos \left(\pi - \phi\right) = -\cos \left(\phi\right), \qquad \phi_{\rm RS} = \pi - 2\phi,
\end{equation}
we assemble all the contributions to determine the total weak deflection angle via the Gauss-Bonnet formalism, incorporating both metric deformation and finite-distance corrections:
\begin{align}
    \alpha &\sim \frac{2 M  \left(v^{2}+1\right) \sqrt{1-b^{2} u^{2}}}{v^{2} b} \nonumber \\
    &-\frac{\varepsilon M^2  \left[\pi -2 \arcsin \! \left(b u \right)\right](v^2+2)}{4 b^{2}v^2}  \nonumber \\
    &- \frac{ M \varepsilon^2  \left(v^{2}+1\right) \sqrt{1-b^{2} u^{2}}}{8v^{2} b} - \mathcal{O}(\varepsilon^3)
\end{align}
when $u \to 0$,
\begin{align}
    \alpha^{\rm massive} &\sim \frac{2 \left(v^{2}+1\right) M }{v^{2} b}-\frac{ \pi \varepsilon M^2  \left(v^{2}+2\right)}{4 b^{2} v^{2}} \nonumber \\
    & - \frac{M \varepsilon^2 \left(v^{2}+1\right)}{8 b \,v^{2}} - \mathcal{O}(\varepsilon^3)
\end{align}
For photons $(v=1)$,
\begin{equation} \label{wda_null}
    \alpha^{\rm photon} \sim \frac{4 M}{b}-\frac{3\pi M^2\varepsilon}{4 b^{2}} - \frac{M \varepsilon^2}{4b} -  \mathcal{O}(\varepsilon^3),
\end{equation}
which elegantly recovers the classic Einstein result $ \alpha = 4M/b $ in the limit $ \varepsilon \to 0 $, while clearly showing how higher-order quantum corrections manifest as suppressed terms in powers of $ M/b $. These corrections are particularly important in precision lensing observations near the Sun or compact objects, where even sub-arcsecond deviations may be measurable.

Let us find a constraint on $\varepsilon$ using the solar system test. Under the  parametrized post-Newtonian (PPN) framework, the angular deflection of starlight that grazes the Sun is expressed as \cite{Chen:2023bao}
\begin{equation} \label{PPN}
\Theta^{\rm PPN} \simeq \frac{4M_\odot}{R_\odot} \left( \frac{n \pm \Delta}{2} \right),
\end{equation}
where $ \Delta_{PPN} = 0.0003 $  quantifies the uncertainty in the curvature caused by the Sun's immense gravitational field, $n = 1.9998$ \cite{Fomalont_2009}, $M_\odot = 1477 \text{ m}$, and $R_\odot = 6.963 \times 10^{8} \text{ m}$. Comparing Eq. \eqref{PPN} with Eq. \eqref{wda_null}), one can extract some numerical constraint on $ \varepsilon $, which we found as $0.061$ for $\Delta_{PPN} < 0$ (the positive value produces an imaginary result for $\varepsilon$). This is consistent with Solar System observations, while remaining within the theoretically permitted regime where $ \varepsilon < 1 $. This is a crucial point, as overly large values of $ \varepsilon $ would either violate observational bounds or render the metric ill-defined (e.g., due to complex-valued square roots). Notably, only the case $ \Delta_{PPN} < 0 $ yields a real, physically meaningful value for $ \varepsilon $, suggesting that the best-fit correction is slightly reducing the deflection angle compared to GR. This finding aligns with your shadow-based analysis, where moderate positive values of $ \varepsilon $ were also shown to be compatible with observations. Furthermore, the second-order $ \varepsilon^2 $ term, while subleading, contributes a finite correction that becomes non-negligible for high-precision measurements and may play a role in future solar-system or pulsar-lensing tests.

\section{Hawking Temperature  of the BHGCE}
\label{sec:hawking-temperature}

Black hole thermodynamics bridges classical gravity, quantum field theory, and holographic principles, thus providing profound insights into quantum gravity. The concept of temperature, especially Hawking and Unruh temperatures, serves as a fundamental cornerstone in understanding these connections. In this context, Hawking radiation emerges naturally as a consequence of quantum effects near black hole horizons, and deviations from classical General Relativity (GR) could manifest distinctively in the thermodynamic properties of black holes, potentially observable through astrophysical phenomena.

To determine the local acceleration and temperature experienced by stationary observers, it is instructive to introduce a generalized gravitational potential, defined through the timelike Killing vector field \(\xi_\alpha\):
\begin{equation}
\phi = \frac{1}{2}\log\left(-g^{\alpha\beta}\xi_\alpha\xi_\beta\right),
\end{equation}
where the normalization condition \(\phi\rightarrow 0\) as \(r\rightarrow\infty\) ensures an asymptotically flat spacetime.

The local gravitational acceleration, \(a^\alpha\), experienced by static observers is then given by:
\begin{equation}
a^\alpha = -g^{\alpha\beta}\nabla_\beta \phi,
\end{equation}
expressing the covariant gradient of the gravitational potential in curved spacetime.

Consequently, the temperature experienced locally by stationary observers (the Unruh-Verlinde temperature) can be written as \cite{Verlinde:2010hp}:
\begin{equation}
T = \frac{\hbar}{2\pi}n^\alpha e^\phi \nabla_\alpha \phi,
\end{equation}
with \(n^\alpha\) representing the unit normal to the holographic screen at radius \(r\), and \(e^\phi\) the gravitational redshift factor.

We now specialize to the line element describing a static and spherically symmetric spacetime,
\begin{equation}
ds^2 = -A(r)dt^2 + B(r)dr^2 + r^2(d\theta^2+\sin^2\theta\,d\phi^2)\,,
\end{equation}
where, in our scenario, \(A(r)=f(r)\) and \(B(r)=1/f(r)\).

The relevant timelike Killing vector, reflecting the stationary symmetry of spacetime, is:
\begin{equation}
\xi_\alpha=(-A(r),\,0,\,0,\,0)\,.
\end{equation}
Thus, the gravitational potential simplifies to:
\begin{equation}
\phi=\frac{1}{2}\log[A(r)]\,.
\end{equation}

From this potential, the radial gravitational acceleration becomes explicitly:
\begin{equation}
a^r=\frac{A'(r)}{2A(r)B(r)}=\frac{A'(r)f(r)}{2A(r)}=\frac{f'(r)}{2}\,.
\end{equation}

Then, the Unruh-Verlinde temperature for our modified spacetime can be succinctly expressed as:
\begin{equation}
T=\frac{\hbar}{4\pi}\frac{A'(r)}{\sqrt{A(r)B(r)}}=\frac{\hbar}{4\pi}f'(r)\,.
\label{eq:unruh-temperature}
\end{equation}

Evaluating explicitly, we have:
\begin{equation}
f'(r)=\frac{2M}{r^2}-\frac{2\varepsilon M^2}{r^3},
\end{equation}
thus yielding the generalized Hawking temperature at radius \(r\):
\begin{equation}
T(r)=\frac{\hbar}{2\pi}\left(\frac{M}{r^2}-\frac{\varepsilon M^2}{r^3}\right).
\end{equation}

The horizon radius, defined by \(f(r_H)=0\), is found as:
\begin{equation}
r_H=M\Sigma\left(1+\sqrt{1-\varepsilon}\right).
\end{equation}
This explicitly demonstrates how the horizon structure itself depends critically on \(\varepsilon\), influencing observable thermodynamics.

The corresponding Hawking temperature at the horizon, \(T_H=T(r_H)\), explicitly becomes:
\begin{equation}
T_H=\frac{\hbar}{4\pi}\frac{f'(r_H)}{\sqrt{f(r_H)f^{-1}(r_H)}}=\frac{\hbar}{4\pi}f'(r_H).
\end{equation}
Substituting explicitly, we find:
\begin{equation}
T_H=\frac{\hbar}{2\pi}\frac{\sqrt{1-\varepsilon}}{M\Sigma\left(1+\sqrt{1-\varepsilon}\right)^2}.
\label{eq:hawking-temp-final}
\end{equation}

The result in Eq.~\eqref{eq:hawking-temp-final} provides a physically transparent interpretation: the parameter \(\varepsilon\), representing quantum or cosmological corrections, significantly modifies the horizon structure and thus the thermal properties of the black hole. Specifically, the temperature is lowered by positive \(\varepsilon\) values, suggesting reduced quantum evaporation rates. Conversely, negative values could enhance Hawking radiation. Such effects are critical in scenarios of primordial black holes, where evaporation timescales and resultant gravitational wave signatures are highly sensitive to small modifications in the black hole temperature.

From an observational perspective, these corrections become increasingly relevant as gravitational-wave astronomy matures. Deviations from classical Hawking radiation predictions, potentially measurable indirectly via black hole population statistics or directly via stochastic gravitational wave backgrounds, could place stringent constraints on \(\varepsilon\) and \(\Sigma\). This scenario underscores the broader scientific value of exploring Hawking temperatures within modified spacetimes, connecting quantum gravity hypotheses with astrophysical observables.

Moreover, the formulation presented here elegantly connects thermodynamics with the dynamics of null geodesics, since the surface gravity—and thus the Hawking temperature—is closely tied to the photon sphere properties. Indeed, as we examine below, the location of the photon sphere, angular velocity, and associated quasinormal mode frequencies are directly influenced by the same parameters (\(\varepsilon\), \(\Sigma\)) affecting temperature. Consequently, combined observational signatures from black hole shadows, gravitational lensing, and gravitational-wave ringdown phases offer robust, complementary methods to test fundamental theories beyond GR.


\section{Eikonal Quasinormal Mode Frequencies of BHGCE} \label{sec7}

Quasinormal modes (QNMs) represent characteristic oscillations of black holes, arising from perturbations in spacetime geometry that propagate outward and gradually decay due to gravitational radiation. The precise determination of QNM frequencies provides an essential observational test of General Relativity (GR) and its potential modifications. In particular, the eikonal (large angular momentum, \(l\gg1\)) limit of QNMs has a profound physical interpretation: it directly connects the oscillation frequencies to the properties of the unstable circular photon orbits (the photon sphere), thereby offering a clear astrophysical signature of deviations from GR.

In the eikonal limit, scalar, electromagnetic, or gravitational perturbations around black holes can be approximated by an effective potential of the form:
\begin{equation}
V_{\rm eff}(r)\simeq f(r)\frac{l^2}{r^2},
\label{eq:Veff}
\end{equation}
with the photon sphere defined by the maximum of the function:
\begin{equation}
g(r)=\frac{f(r)}{r^2}=\frac{1}{r^2}-\frac{2M\Sigma}{r^3}+\frac{\varepsilon M^2\Sigma^2}{r^4}.
\label{eq:gr}
\end{equation}

The photon sphere radius, \(r_0\), corresponds to the unstable circular orbit radius of photons, determined by the condition \(g'(r_0)=0\). Explicit calculation yields:
\begin{equation}
r_0=M\Sigma\left(\frac{3+\sqrt{9-8\varepsilon}}{2}\right).
\label{eq:r0}
\end{equation}
In the Schwarzschild limit (\(\Sigma=1,\varepsilon=0\)), Eq.~\eqref{eq:r0} recovers the well-known photon sphere radius \(r_0=3M\) \cite{Claudel:2000yi}. The deviation parameters \(\Sigma\) and \(\varepsilon\) thus directly influence the position of the photon sphere, shifting it inward or outward from the standard GR value.

Physically, a shift in the photon sphere radius modifies key observational features, including black hole shadows observed by very-long-baseline interferometry (e.g., Event Horizon Telescope observations) and gravitational lensing patterns. Precise measurements of these observables provide constraints on \(\Sigma\) and \(\varepsilon\), probing theories beyond standard GR predictions.

The angular velocity of null geodesics orbiting at the photon sphere, \(\Omega_0\), is crucial since it determines the characteristic frequency of the oscillatory component of the quasinormal modes. It is given by:
\begin{equation}
\Omega_0=\sqrt{\frac{f(r_0)}{r_0^2}}.
\end{equation}

Evaluating explicitly with the photon sphere radius \(r_0\) defined in Eq.~\eqref{eq:r0}, we find:
\begin{equation}
\Omega_0=\frac{\sqrt{\alpha^2-2\alpha+\varepsilon}}{M\Sigma\,\alpha^2},\quad \text{where}\quad \alpha\equiv\frac{3+\sqrt{9-8\varepsilon}}{2}.
\label{eq:Omega0_final}
\end{equation}

This expression makes explicit how deviations from GR encoded in \(\Sigma\) and \(\varepsilon\) alter photon sphere geodesic motion. A change in \(\Omega_0\) implies modifications to observed gravitational lensing ring radii and angular positions of photon orbits, potentially detectable with high-precision gravitational lensing measurements.

Another critical quantity is the Lyapunov exponent, \(\lambda\), characterizing the timescale for instability of the photon orbit. Physically, \(\lambda\) determines how rapidly perturbations diverge from the photon sphere orbit, controlling the damping rate of the associated QNMs. It is defined as:
\begin{equation}
\lambda=r_0\sqrt{-\frac{f(r_0)}{2}\left.\frac{d^2}{dr^2}\frac{f(r)}{r^2}\right|_{r=r_0}}.
\end{equation}

Explicit computation gives:
\begin{equation}
\lambda=\frac{\alpha}{M\Sigma}\sqrt{-\frac{1}{2}\left(1-\frac{2}{\alpha}+\frac{\varepsilon}{\alpha^2}\right)\left(\frac{6}{\alpha^4}-\frac{24}{\alpha^5}+\frac{20\varepsilon}{\alpha^6}\right)}.
\label{eq:lambda_final}
\end{equation}

The Lyapunov exponent sensitively depends on the modification parameters \(\Sigma\) and \(\varepsilon\). Physically, a larger \(\lambda\) indicates faster damping of the perturbation, thus potentially altering the duration and visibility of gravitational wave ringdown signals observed by detectors like LIGO, Virgo, and the upcoming Einstein Telescope. Precise ringdown observations may therefore strongly constrain modifications to the black hole metric.

In the eikonal regime, QNM frequencies (\(\omega\)) exhibit a direct relationship to photon sphere characteristics:
\begin{equation}
\omega=l\,\Omega_0 - i\left(n+\frac{1}{2}\right)\lambda,
\label{eq:omega_eikonal}
\end{equation}
where \(l\gg1\) is the angular mode number and \(n\) the overtone number. Using the expressions from Eqs.~\eqref{eq:Omega0_final} and \eqref{eq:lambda_final}, the QNM spectrum explicitly reveals how \(\Sigma\) and \(\varepsilon\) influence both oscillatory and damping aspects of perturbations.

Astrophysically, this result has significant implications. The real part, governed by \(\Omega_0\), sets the frequency of gravitational wave oscillations during black hole mergers, while the imaginary part, controlled by \(\lambda\), governs the damping and decay time. Deviations from Schwarzschild predictions, characterized by \(\Sigma\neq1\) or \(\varepsilon\neq0\), could manifest clearly in observed gravitational waveforms, especially during the post-merger ringdown phase. Such signals provide stringent tests of gravitational theories, placing robust constraints on possible modifications arising from quantum gravity corrections, extra dimensions, scalar fields, or dark-sector physics.

Our detailed derivation and explicit analysis demonstrate how the geometry surrounding black holes, modified through the parameters \(\Sigma\) and \(\varepsilon\), directly impacts photon sphere properties and consequently, quasinormal mode spectra. High precision measurements from current and next generation gravitational wave observatories, combined with electromagnetic observations from black hole shadow imaging, offer promising avenues to observationally constrain these parameters. Thus, black hole spectroscopy carefully analyzing the QNM frequencies and damping rates emerges as a powerful astrophysical tool to explore and potentially validate theories of gravity beyond General Relativity.

\section{Gravitational Redshift of the BHGCE} \label{sec8}

When a light pulse travels through a gravitational field, its frequency and photon energy decrease between the emission and reception events, a phenomenon referred to as gravitational redshift. Mathematically, this redshift \(z\) is defined by \cite{Misner:1973prb,Nicolini:2009gw,Cardenas:2021eri}
\begin{equation}
1 + z \;=\; \frac{\omega_{e}}{\omega_{o}},
\end{equation}
where \(\omega\) is the frequency, and the subscripts \(e\) and \(o\) denote the emitter and the observer, respectively.

For a static observer in a spacetime possessing a timelike Killing vector field, \( \xi^{\mu} = (\partial_t)^\mu \), one has
\begin{equation}
\xi^\mu \xi_\mu = g_{tt} = -f(r).
\end{equation}
A static observer at a constant radial coordinate \( r \) has a four-velocity
\begin{equation}
u^\mu = \frac{1}{\sqrt{-g_{tt}}}\,(1,0,0,0) = \left(\frac{1}{\sqrt{f(r)}}, 0,0,0\right).
\end{equation}
Let \( k^\mu \) denote the four-momentum of a photon. In the presence of a Killing vector field, the conserved quantity
\begin{equation}
E = -\xi^\mu k_\mu,
\end{equation}
remains constant along the photon trajectory. The frequency measured by an observer with four velocity \( u^\mu \) is given by
\begin{equation}
\nu = - k_\mu u^\mu.
\end{equation}
Thus, at the emission point \( r=r_e \) and at the observation point (taken to be \( r \to \infty \), where \( g_{tt}(\infty)\to -1 \)), one obtains 
\begin{equation}
\nu_e \sqrt{f(r_e)} = \nu_\infty.
\end{equation}
Here, \( \nu_e \) and \( \nu_\infty \) are the photon frequencies measured by the static observer at \( r_e \) and at infinity, respectively.

Defining the gravitational redshift \( z \) as
\begin{equation}
1+z \equiv \frac{\nu_e}{\nu_\infty},
\end{equation}
we arrive at
\begin{equation}
1+z = \frac{1}{\sqrt{f(r_e)}}.
\end{equation}
Substituting Eq. (2) into Eq. (9) yields the final form:
\begin{equation} \label{redshift1}
z = \frac{1}{\sqrt{1 - \frac{2M\Sigma}{r_e} + \frac{\varepsilon M^2\Sigma^2}{r_e^2}}} - 1.
\end{equation}

\begin{figure}[ht!]
\centering
\includegraphics[scale=0.6]{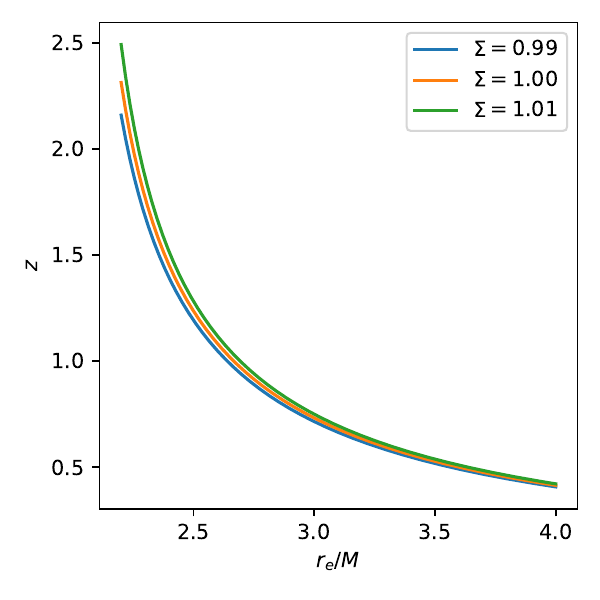} \
\includegraphics[scale=0.6]{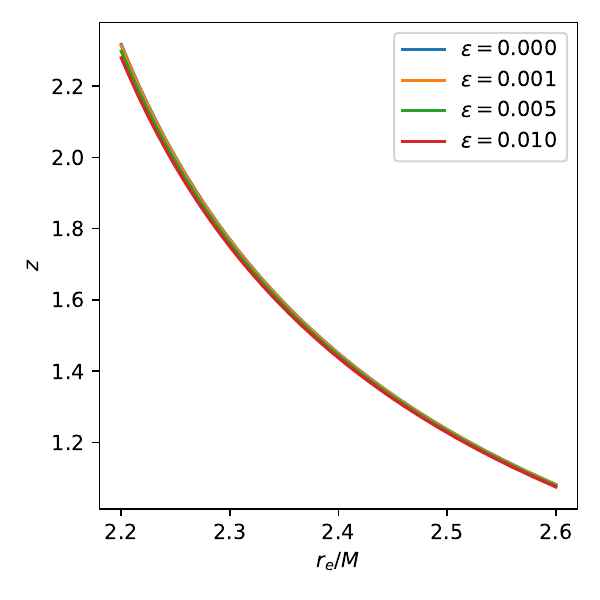} \
\caption{
Gravitational redshift z versus $r_e/M$ }
\label{fig:1} 	
\end{figure}

Equation encapsulates the gravitational redshift in the RN metric and is shown in Fig. \ref{fig:1}. Notice that in the limit \( \varepsilon  \to 0 \) and \( \Sigma  \to 1 \), Eq. \ref{redshift1}  reduces to the well-known redshift formula for the Schwarzschild black hole,
\begin{equation}
  z = \frac{1}{\sqrt{1 - \frac{2M}{r_e}}} - 1.
\end{equation}
The domain of validity of the event horizon is for \( r_e>r_+ \), where \( r_+ \) is the outer (event) horizon. As \( r_e \to r_+ \) the function \( f(r_e) \) approaches zero, leading to \( z\to \infty \) which reflects the infinite redshift at the horizon.

In summary, gravitational redshift encapsulates how the energy of a photon diminishes when traveling from an emitter deeper in a gravitational well to an observer located farther out (or even at spatial infinity). The magnitude of \(z\) depends on the spacetime geometry, the motion of the emitter (e.g., circular orbits), and the observer’s location and velocity.

\section{Ringdown Waveform of the BHGCE} \label{sec9}

In our study, we restrict attention to static, spherically symmetric black hole geometries. The line element is expressed in its most general form as
\begin{eqnarray}
ds^2 = -A(r)\,dt^2 + \frac{dr^2}{B(r)} + r^2\,d\Omega_2^2,
\end{eqnarray}
with the outer event horizon located at \(r=r_+\) where \(A(r_+)=B(r_+)=0\).

To isolate the effect of the metric on black hole echoes, we perturb the spacetime by a free, massless scalar field \(\psi(t,r,\theta,\phi)\), which obeys
\begin{eqnarray}
\Box\psi \equiv \frac{1}{\sqrt{-g}}\,\partial_\mu\Bigl(\sqrt{-g}\,g^{\mu\nu}\partial_\nu\psi\Bigr)=0.
\end{eqnarray}
Employing the standard separation of variables,
\begin{eqnarray}
\psi(t,r,\theta,\phi)=\sum_{l,m}\frac{\Phi(t,r)}{r}\,Y_{l,m}(\theta,\phi),
\end{eqnarray}
reduces the field equation to a radial and temporal differential equation:
\begin{eqnarray}
-\frac{\partial^2\Phi(t,r)}{\partial t^2} + A\,B\,\frac{\partial^2\Phi(t,r)}{\partial r^2} + \frac{1}{2}\Bigl(B\,A' + A\,B'\Bigr)\frac{\partial\Phi(t,r)}{\partial r}\\ - \left[\frac{l(l+1)}{r^2}\,A + \frac{1}{2r}\Bigl(AB\Bigr)'\right]\Phi(t,r)=0 \notag,
\end{eqnarray}
where the prime indicates differentiation with respect to \(r\); the coordinate \(r\) ranges over \((r_+, \infty)\).

\begin{figure}[!h]
   \includegraphics[scale = 0.4]{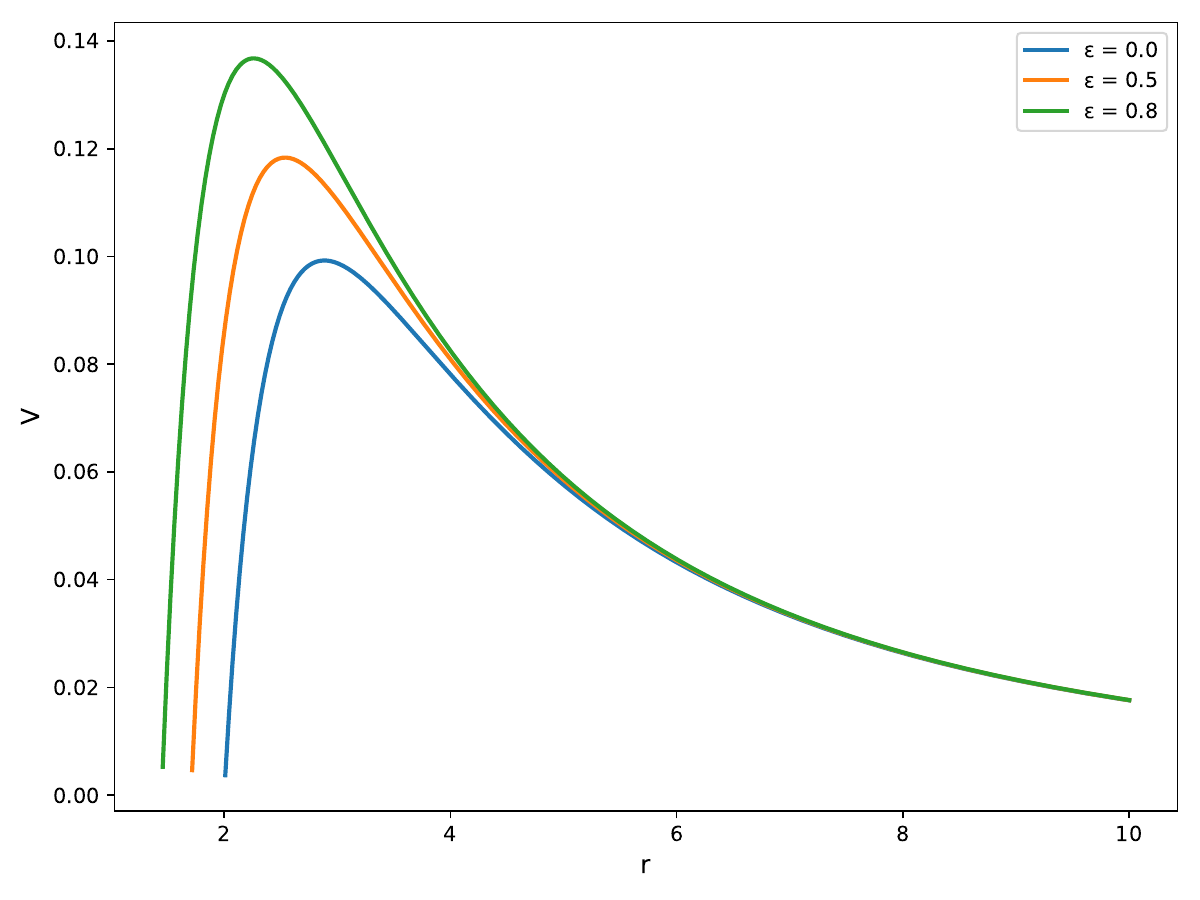}
   \includegraphics[scale = 0.4]{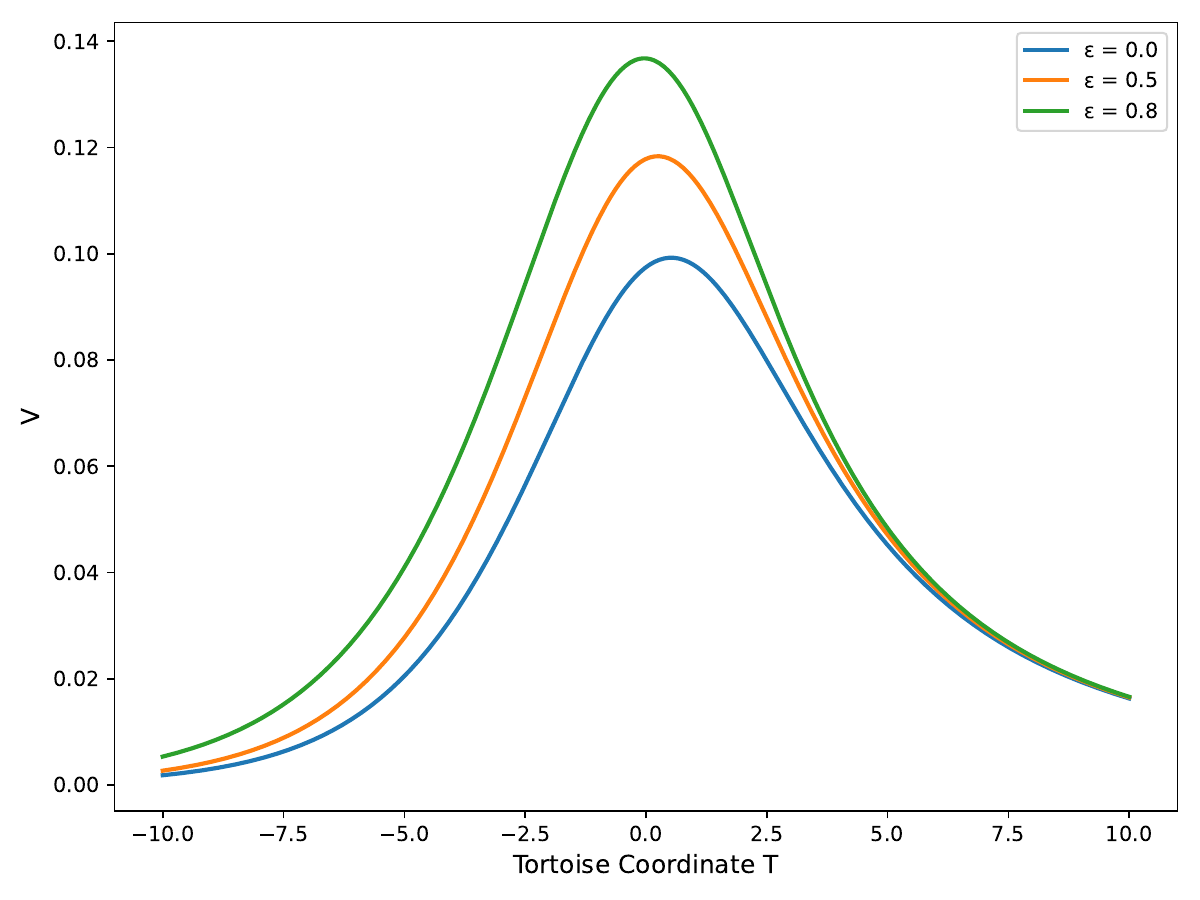}
\caption{Variation of the scalar potential $V_0(r)$ with the radial distance $r$ for different values of the parameter $\varepsilon$ with $M=1$, $\Sigma=1$. }
\label{fig_Vs_01}
\end{figure}

For quasinormal mode (QNM) calculations, the appropriate boundary conditions are imposed: the field is purely ingoing at the horizon (\(r\to r_+\)) and purely outgoing at infinity. To facilitate this analysis, we map the radial coordinate to the tortoise coordinate \(r_*\) defined by
\begin{eqnarray}
dr_*=\frac{dr}{\sqrt{A\,B}},
\end{eqnarray}
which transforms the wave equation into
\begin{eqnarray}
-\frac{\partial^2\Phi}{\partial t^2} + \frac{\partial^2\Phi}{\partial r_*^2} - V(r)\,\Phi = 0,
\end{eqnarray}
with the effective potential given by
\begin{eqnarray}
V_0(r) = \frac{l(l+1)}{r^2}\,A + \frac{1}{2r}\Bigl(AB\Bigr)'.
\end{eqnarray}
The transformation to \(r_*\) reparameterizes the spatial domain (stretching it to \((-\infty,\infty)\)) while preserving the intrinsic structure of the potential, particularly the number of its extrema. In upper Figure \ref{fig_Vs_01} we plot the radial profile of the effective potential \(V(r)\) for three representative values of the deformation parameter, \(\varepsilon=0.0,\;0.5,\;0.8\). As \(\varepsilon\) increases, the height of the potential barrier grows monotonically, signaling a stronger trapping region for perturbations. Moreover, the location of the peak shifts slightly outward (toward larger \(r\)) with larger \(\varepsilon\), indicating that the photon‑sphere radius is pushed to larger radii by the quantum‑vacuum backreaction. Lower Figure \ref{fig_Vs_01} shows the same potential plotted against the tortoise coordinate \(r_*\), which logarithmically stretches the near–horizon region. Here, one sees that the barrier becomes both taller and narrower in tortoise space as \(\varepsilon\) increases: the uphill “walls’’ on either side of the peak steepen, reflecting the fact that modes take longer (in coordinate time) to traverse the potential as quantum corrections strengthen.

A brief analysis shows that \(V(r_+)=0\) with a positive slope, \(V'(r_+)>0\). At spatial infinity, assuming an asymptotically flat form (\(h\sim f\sim 1-2M/r+\cdots\)), one obtains \(V(\infty)\rightarrow0^+\) and \(V'(\infty)\rightarrow0^-\). Consequently, the effective potential must exhibit at least one maximum. In standard black holes such as Schwarzschild or Reissner-Nordstr\"om, the single-peak structure precludes echo phenomena, which require a resonant cavity typically associated with at least three extrema—a minimum bounded by two maxima. Notably, for large angular momentum \(l\) the dominant term behaves as \(V\sim h/r^2\), corresponding to the photon sphere; thus, multiple peaks in \(V(r)\) are closely linked to the presence of multiple photon spheres, a feature rarely encountered in classical black hole solutions.

To extract the time-domain evolution of \(\Phi(t,r)\) as governed by the wave equation, we employ a finite difference approach a method that has proven effective in probing QNMs for both black hole \cite{Zhu:2014sya} and wormhole \cite{Liu:2020qia} geometries.

The temporal and spatial variables are discretized by setting \(t=i\,\Delta t\) and \(r_*=j\,\Delta r_*\) (with \(i,j\in\mathbb{Z}\)), and the scalar field and potential are correspondingly denoted by
\begin{eqnarray}
\Phi(t,r_*)\equiv\Phi_{i,j}, \qquad V(r_*)\equiv V_j.
\end{eqnarray}
Under these definitions, the wave equation is approximated by the finite difference equation:
\begin{eqnarray}
-\frac{\Phi_{i+1,j}-2\Phi_{i,j}+\Phi_{i-1,j}}{\Delta t^2} + \frac{\Phi_{i,j+1}-2\Phi_{i,j}+\Phi_{i,j-1}}{\Delta r_*^2} \notag\\ - V_j\,\Phi_{i,j}=0.
\end{eqnarray}

For the initial condition we adopt a Gaussian wave packet:
\begin{eqnarray}
\psi(t=0,r_*)=\exp\left[-\frac{(r_*-\bar{a})^2}{2}\right],\qquad\psi(t<0,r_*)=0,
\end{eqnarray}
where \(\bar{a}\) defines the center of the packet. We explore two distinct configurations: when the packet is situated outside the double-peaked region of the potential, and when it is located within the well between the peaks. In the latter scenario, as the initial packet is placed nearer to the potential minimum, echo signals become more pronounced and the characteristic echo frequency approximately doubles relative to the exterior configuration.

Boundary conditions are imposed by requiring
\begin{eqnarray}
\Phi(t,r_*)\bigg|_{r_*\to -\infty}=e^{-i\omega r_*},\qquad \Phi(t,r_*)\bigg|_{r_*\to +\infty}=e^{i\omega r_*},
\end{eqnarray}
with \(\omega\) denoting the QNM frequency (distinct from the echo frequency). The update rule is then formulated as
\begin{eqnarray}
\Phi_{i+1,j} = -\Phi_{i-1,j} + \left[2 - 2\frac{\Delta t^2}{\Delta r_*^2} - \Delta t^2\,V_j\right]\Phi_{i,j}\\ \notag + \frac{\Delta t^2}{\Delta r_*^2}\Bigl(\Phi_{i,j+1}+\Phi_{i,j-1}\Bigr).
\end{eqnarray}
To ensure numerical stability, the time step is chosen in compliance with the von Neumann criterion, specifically \(\Delta t/\Delta r_*=0.5\) (see, e.g., \cite{Zhu:2014sya} for further details).

\begin{figure}[htbp]
   \includegraphics[scale = 0.4]{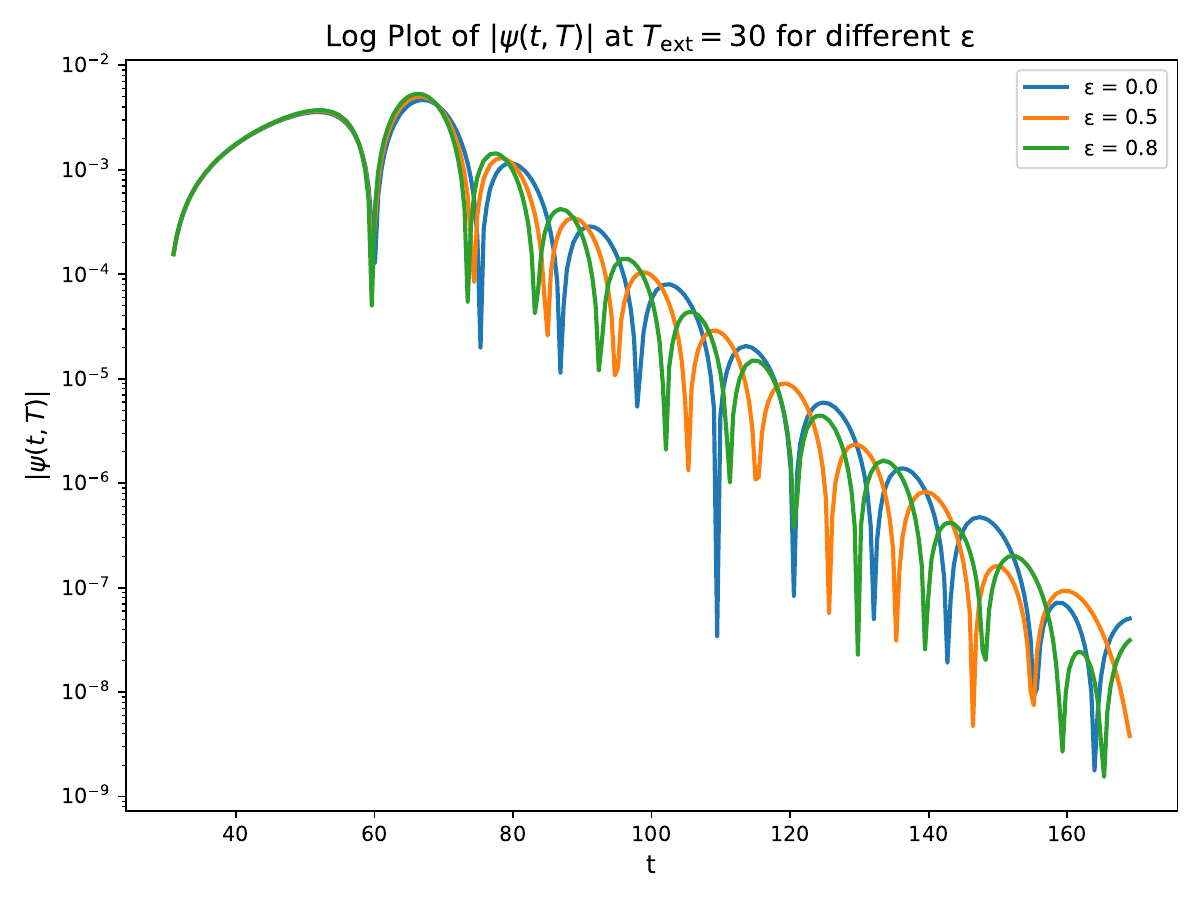}
   \includegraphics[scale = 0.4]{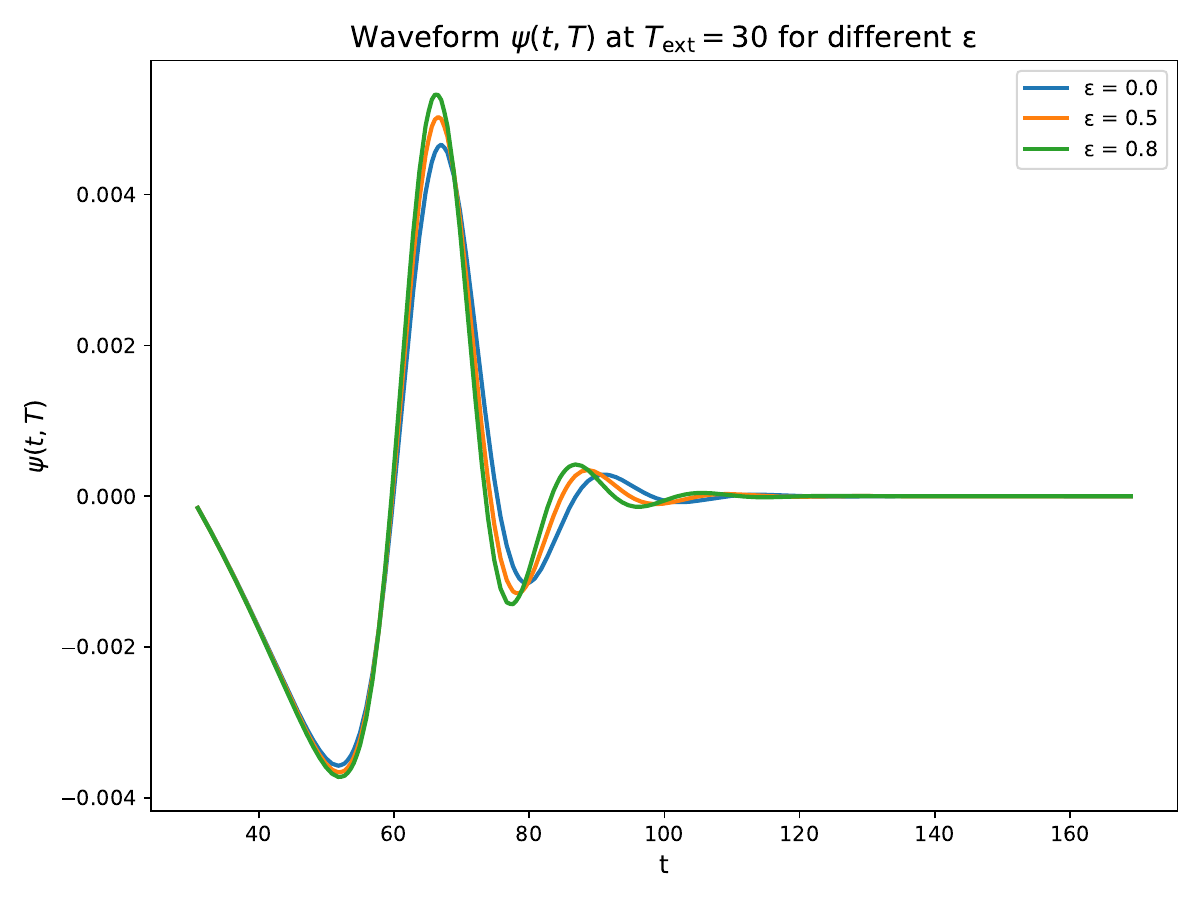} 
\caption{The time-domain profiles of the massless scalar
perturbations for different multipole moments $l$ with the parameter values $\varepsilon$, $G = 1$, $M= 1$, $\Sigma =1$.}
\label{time01}
\end{figure}

In the lower panel of Figure \ref{time01} we display the late‑time ringdown signal \(\Psi(t)\) (extracted at a fixed observation radius) for the same set of \(\varepsilon\) values. With increasing \(\varepsilon\), the oscillation period visibly lengthens, and the envelope decays more slowly. In the upper panel of Figure \ref{time01} we plot \(\ln|\Psi(t)|\), which makes the quasi‑periodic oscillations and exponential damping far more transparent:
Oscillation Frequency: The spacing between successive peaks in the log‑plot clearly increases with \(\varepsilon\), indicating that the real part of the quasinormal frequency decreases as the GCW deformation grows. Damping Rate: The magnitude of the slope of the peak‑envelope in the \(\ln|\Psi|\) plot (i.e.\ the imaginary part of the mode) diminishes for larger \(\varepsilon\), showing that the ringdown damps more slowly under stronger quantum‑vacuum corrections.

Together, these results demonstrate that both the oscillation frequency and damping rate of the dominant quasinormal mode are sensitive probes of the generalized Compton wavelength parameter.

\section{Conclusion} \label{conc}
In this work, we have investigated the implications of a generalized Compton wavelength framework, embedded in a three-dimensional dynamical quantum vacuum (3D DQV), on a Schwarzschild-like black hole geometry. The introduction of the quantum deformation parameter $ \varepsilon $, which encapsulates backreaction effects due to vacuum energy density fluctuations, modifies the black hole metric in a controlled and theoretically consistent manner. Through analytical methods, we derived explicit corrections to key physical observables and assessed their phenomenological viability.

The black hole shadow analysis in Section III provided an exact expression for the shadow radius, yielding bounds on $ \varepsilon $ from EHT observations of Sgr. A* and M87*. These constraints lie within the domain of theoretical consistency and do not exclude significant deviations from general relativity, particularly in the case of Sgr. A*, where negative $ \varepsilon $ values corresponding to enhanced gravitational attraction remain observationally viable. The weak deflection angle was computed via two independent methods: the Keeton-Petters post-post-Newtonian expansion and the Gauss-Bonnet topological approach. Both revealed non-trivial corrections scaling with $ \varepsilon $ and confirmed that solar system lensing observations restrict $ \varepsilon $ to values near zero, with a representative upper bound of $ \varepsilon \approx 0.061 $ consistent with PPN constraints.

Thermodynamically, we found that the Hawking temperature of the modified black hole depends nonlinearly on $ \varepsilon $, with positive deformation reducing the evaporation rate—a feature with potential implications for the fate of primordial black holes. Complementing this, the eikonal quasinormal mode frequencies, computed in Section VII, demonstrated that both the real and imaginary parts of the QNM spectrum are sensitive to $ \varepsilon $ and the rescaling factor $ \Sigma $, suggesting that gravitational wave ringdown observations can serve as a sharp probe of quantum corrections to the metric.

We also derived the gravitational redshift and ringdown waveform in the presence of the GCW-induced deformation. The redshift formula remains analytically tractable and reveals that the modification becomes significant near the horizon. Based on scalar perturbations, the ringdown analysis showed how the effective potential, and consequently the waveform morphology, is altered by varying $ \varepsilon $, offering a potential observational signature in the form of modified echo dynamics.

Taken together, these results highlight the observational richness of the GCW-modified black hole metric. With analytical control over corrections to classical phenomena and consistency with existing data, this framework presents a promising semiclassical window into quantum gravity phenomenology. Further constraints may be extracted through multi-messenger observations involving VLBI shadow imaging, gravitational lensing, and gravitational wave spectroscopy, placing tighter bounds on the deformation parameter $ \varepsilon $ and its role in black hole physics.

A natural extension of this work would be exploring the GCW-modified metric's rotating analogs, particularly examining how the generalized Compton wavelength formalism deforms the Kerr geometry. This would enable the analysis of frame dragging, ergosphere structure, and shadow asymmetry, all key signatures in upcoming high-resolution VLBI measurements. Furthermore, coupling this framework to quantum field dynamics in curved spacetime may yield corrections to greybody factors and emission spectra, deepening the connection between semiclassical gravity and observational signatures.

\acknowledgments
R. P.,  A. \"O. and G. L. would like to acknowledge networking support of the COST Action CA21106 - COSMIC WISPers in the Dark Universe: Theory, astrophysics and experiments (CosmicWISPers), the COST Action CA22113 - Fundamental challenges in theoretical physics (THEORY-CHALLENGES), the COST Action CA21136 - Addressing observational tensions in cosmology with systematics and fundamental physics (CosmoVerse), the COST Action CA23130 - Bridging high and low energies in search of quantum gravity (BridgeQG), and the COST Action CA23115 - Relativistic Quantum Information (RQI) funded by COST (European Cooperation in Science and
Technology). A. \"O. also thanks to EMU, TUBITAK, ULAKBIM (Turkiye) and SCOAP3 (Switzerland) for their support.  

\bibliography{ref}

\end{document}